# LLM-Driven Rubric-Based Assessment of Algebraic Competence in Multi-Stage Block Coding Tasks with Design and Field Evaluation

Yong Oh Lee[a], Byeonghun Bang[b], Sejun Oh[c]

[a]*Department of Industrial and Data Engineering, Hongik University, Seoul, 04066, South Korea*
[b]*Department of Computer Engineering, Hongik University, Seoul, 04066, South Korea*
[c]*Department of Mathematics Education, Hongik University, Seoul, 04066, South Korea*

**Abstract**

As online education platforms continue to expand, there is a growing need for assessment methods that not only measure answer accuracy but also capture the depth of students' cognitive processes in alignment with curriculum objectives. This study proposes and evaluates a rubric-based assessment framework powered by a large language model (LLM) for measuring algebraic competence, real-world–context block coding tasks. The problem set, designed by mathematics education experts, aligns each problem segment with five predefined rubric dimensions, enabling the LLM to assess both correctness and quality of students' problem-solving processes. The system was implemented on an online platform that records all intermediate responses and employs the LLM for rubric-aligned achievement evaluation. To examine the practical effectiveness of the proposed framework, we conducted a field study involving 42 middle school students engaged in multi-stage quadratic equation tasks with block coding. The study integrated learner self-assessments and expert ratings to benchmark the system's outputs. The LLM-based rubric evaluation showed strong agreement with expert judgments and consistently produced rubric-aligned, process-oriented feedback. These results demonstrate both the validity and scalability of incorporating LLM-driven rubric assessment into online mathematics and STEM education platforms.

# 1. Introduction

Academic achievement refers to the extent to which learners attain intended learning outcomes across knowledge, skills, and attitudes (Bhat and Bhardwaj (2014)). In mathematics, particularly in the domain of algebra, academic achievement is reflected not only in the correctness of final answers but also in the quality of the problem-solving process(Sinaga et al. (2023)). This includes reasoning, representation, and communication, all of which are essential for identifying misconceptions, tracking learning progress, and designing targeted instructional interventions(Palinussa et al. (2021); Kabadaş and Mumcu (2024)).

A rubric is an assessment tool that specifies explicit criteria and clearly defined performance levels for a given task(Panadero et al. (2023)). When implemented systematically, rubric-based assessment enhances objectivity, transparency, and consistency in evaluation while clarifying performance expectations in alignment with curricular goals. In mathematics, learning tasks often require the integration of multiple concepts and solution strategies, which makes it essential to assess not only individual skills but also overall mastery at the end of the learning sequence. This characteristic underscores the importance of rubric-based assessment in summative evaluation, while also supporting formative purposes by guiding structured feedback that addresses specific learning needs(Tashtoush et al. (2025)). In algebraic tasks, it enables evaluators to determine how effectively students translate problems into symbolic expressions, apply logical reasoning to justify each step, and present solutions with clarity(New York State Education Department (2023)). These capabilities cannot be captured by accuracy-based evaluation alone.

Multi-stage problem-solving tasks often take the form of interconnected questions that progress from basic concept understanding to more complex application problems. Each sub-problem can be evaluated with its own rubric to capture specific skills or concepts, yet such stage-level evaluations do not readily support a comprehensive judgment of a learner's overall competence across the entire task sequence. Designing an integrated rubric that holistically reflects performance over multiple stages is challenging even for experienced teachers, as it requires balancing the relative importance of each stage and articulating coherent performance descriptors(Moskal and Leydens (2000); Jonsson and Svingby (2007)).

These difficulties become even more pronounced in large-scale online learning environments, where the volume of students and submissions renders manual rubric construction and evaluation increasingly impractical. Consolidating multi-stage performance into a single, well-defined framework while providing timely and consistent feedback often exceeds the feasible workload for human evaluators. Conventional automated systems, such as rule-based scoring, can efficiently and reliably assess the correctness of individual steps, but they are less effective when student responses vary in reasoning strategies, representations, or intermediate formulations(Safilian et al. (2025)). Such diversity is common in mathematics problem solving, where equivalent solutions may be expressed through different symbolic forms or logical pathways. This variability increases the complexity of assessment and limits the ability of existing systems to provide a holistic, process-oriented evaluation that captures both accuracy and quality. There is therefore a clear need for scalable, semantically aware methods capable of integrating rubric-based judgments(Hellman et al. (2023)).

Recent advances in artificial intelligence have led to the rapid development of large language models (LLMs) with exceptional capabilities in natural language understanding and generation(Guo et al. (2023)). These models can process a wide range of input formats, including text, code, and mathematical notation, making them highly adaptable to diverse educational contexts. In the field of education, LLMs have been increasingly applied to tasks such as automated grading, personalized feedback generation, intelligent tutoring, and the creation of customized learning materials(Sharma et al. (2025)).

Because LLMs can interpret unstructured or semi-structured responses including open-ended explanations, programming code, and algebraic expressions, they are well suited for tasks that require evaluating both the correctness and the reasoning quality of a solution. This capability positions them as a strong candidate for advancing rubric-based academic achievement assessment in mathematics and computational tasks(Morris et al. (2025)).

While progress has been made in harnessing LLMs for automated grading—especially in essay and short-answer domains—systematic research and datasets for multi-stage, rubric-based assessment of mathematical and coding competencies are only beginning to emerge(Fagbohun et al. (2024)). Automated short-answer grading has emerged as an efficient solution for evaluating student responses and has been refined with rubric-aligned strategies to improve reliability and fairness. Much of this research, however, concentrates on single-step answers or writing tasks as well as automated essay scoring(Henkel et al. (2024); Pack et al. (2024); Tang et al. (2024); Xiao et al. (2024)). In the context of mathematics education, recent developments leverage LLMs for marking solutions to algebraic expressions and mathematical reasoning tasks, employing frameworks that automatically verify proofs and compare generated answers with reference solutions. Yet, these evaluations typically measure correctness or procedural quality at the single-step or whole-answer level, rather than systematically dissecting student reasoning across multiple steps(Wang et al. (2025); Fang et al. (2024)). In addition, rubric-based assessments tailored to multi-stage math and coding tasks remain sparse. Some recent work begins to address this gap: for example, pointwise rubric evaluation frameworks propose using LLMs to judge each rubric dimension individually in code or math exercises, providing more granular and process-oriented feedback than holistic scoring alone. However, even in these cases, large publicly available datasets or standardized benchmarks for multi-stage, rubric-aligned math and coding assessment are rare, and most datasets remain proprietary or in development (Pathak et al. (2025)).

This study makes several distinctive contributions to the field of technology-enhanced mathematics education. First, it introduces a rubric–problem segment mapping framework in which domain experts systematically align each step of a problem-solving sequence with predefined rubric dimensions. This design enables an LLM to conduct consistent, criterion-referenced assessments of students' performance, going beyond surface-level correctness to evaluate problem-solving processes in depth. Second, the study develops a real-world–context, stepwise problem set that integrates traditional mathematical reasoning with block-based coding activities. By embedding the problem within an authentic scenario and progressively increasing task complexity from algebraic formulation to computational implementation, this approach fosters both conceptual understanding and computational thinking. Third, the system is implemented on an online learning platform that records all intermediate student responses, applies rule-based scoring for objective correctness, and leverages LLM-generated evaluations for comprehensive achievement assessment. Finally, the effectiveness and practicality of the proposed framework are empirically validated through a field study involving 43 middle school students, incorporating both learner self-assessments and expert evaluations of self-assessment quality. Collectively, these contributions advance the integration of expert-defined rubrics, process-oriented LLM evaluation, and authentic problem design within real classroom environments.

Table 1: Stages of the multi-stage problem-solving task and time allocation

| Phase | Time | Stage | Activity |
|---|---|---|---|
| Introduction | 5 min | #1 | Turning off an alarm with finding two consecutive natural numbers (easy-to-infer target number) |
| | | #2 | Presenting a problem situation using a four-panel comic (linking the alarm-off task to quadratic equations and block coding) |
| Development | 30 min | #3 | Practicing block coding |
| | | #4 | Finding two consecutive natural numbers (easy case) using quadratic equations and block coding |
| | | #5 | Finding two consecutive natural numbers (hard case) using quadratic equations and block coding |
| Conclusion | 5 min | #6 | Explaining the relationship between the value obtained through block coding and quadratic equations |
| | | Self-check | Post-task survey with five Likert items and a one-sentence learning reflection used as auxiliary evidence in the final assessment |

## 2. Methods

*2.1. Problem Design and Implementation*

*2.1.1. Overview of Multi-Stage Problem-Solving Task and Rubric Framework*

The problem was designed as a multi-stage task sequence that progresses from the application of previously learned concepts to real-world mathematical modeling and problem-solving. The instructional context focused on finding two consecutive natural numbers whose product is a given value, formulated as a quadratic equation. The activity was positioned as a performance assessment within a mathematics class where students had already been introduced to the concept of quadratic equations and their solutions.

The multi-stage problem-solving task was structured to progress from introductory activities to development and concluding reflection phases. As summarized in Table 1, the sequence began with two introductory stages designed to situate the problem in a realistic context (#1 Turning off a student's alarm, #2 Extending the alarm-off task to everyone using quadratic equations and block coding). The development phase included a block coding practice stage (#3) followed by two problem-solving stages (#4–#5) that required finding two consecutive natural numbers whose product matched a given target value. #4 used a relatively easy-to-infer target number (110), while #5 employed a more difficult case (8,742) to increase cognitive demand. The final stage (#6) required students to explain their solution to a quadratic equation. After completing all stages, students engaged in a self-check activity involving a five-item Likert-scale survey and a one-sentence learning reflection. These self-reports were incorporated as auxiliary evidence in the final academic achievement evaluation, based on prior findings that self-assessment can enhance metacognitive awareness and support accurate performance appraisal (Köppe et al. (2024)).

The assessment rubric, developed by mathematics education experts, encompassed three major domains: *Knowledge and Understanding*, *Procedural Skills*, and *Values and Attitudes*. Each domain contained performance descriptors for three achievement levels (High, Medium, Low), as detailed in Table 2. The Knowledge and Understanding domain evaluated conceptual mastery of quadratic equations and their solutions. The Procedural Skills domain assessed the ability to represent problem-solving procedures logically and systematically through block coding. The Values and Attitudes domain measured the learner's willingness to tackle challenging real-world problems through mathematical modeling and to use computational tools proactively. This rubric served as the foundation for the subsequent mapping of problem segments to assessment criteria in the academic achievement evaluation phase.

Table 2: Rubric for academic achievement across three domains and performance levels

| Domain | High | Medium | Low |
|---|---|---|---|
| Knowledge and Understanding | Understands quadratic equations and their solution | Checks whether a given number satisfies a quadratic equation | Recognizes solutions only when guided step-by-step |
| Procedural Skills | Solves quadratic equations logically via block coding | Solves with guidance using block coding | Recognizes quadratic equations can be solved with block coding |
| Values and Attitudes | Proactively applies computational tools to real-world problems | Attempts to use computational tools in problem solving | Shows willingness to tackle real-world problems |

*2.1.2. Detailed Task Structure*

The multi-stage problem-solving task incorporated a narrative element at the introduction stage. Specifically, stage #2 was framed as a four-panel comic featuring two fictional students, Jiho and Yujin, who encounter situations that require identifying two consecutive natural numbers whose product equals a given target number. This storyline provided a realistic context and supported student engagement. The development phase then varied the difficulty of target values (e.g., 110 vs. 8742) to scaffold cognitive demand, while maintaining continuity with the introductory narrative.

**Stage #1** introduces an alarm clock scenario where the device can only be turned off by entering two consecutive natural numbers whose product equals a given target number (e.g., 110). Students are guided to: (i) express the larger number in terms of the smaller, (ii) set up the quadratic equation, (iii) solve the equation, and (iv) verify the solution. This stage reinforces basic symbolic manipulation and conceptual understanding of quadratic equations.

**Stage #2** presented as a four-panel comic scenario where two students discuss their experiences. The dialogue contrasts an easy-to-solve problem (finding two consecutive numbers whose product is 110) with a harder-to-infer one (where the product is 8742). The students acknowledge that solving the harder problem with a purely symbolic method like factoring is time-consuming and difficult. This narrative increases the cognitive demand and allows students to experience the limitations of symbolic reasoning firsthand. The stage concludes with the students proposing to use block coding as a more efficient strategy, thus consolidating the need for computational tools which are introduced in the subsequent stages.

**Stage #3** introduces block coding as a scaffold for procedural reasoning. Block coding is a visual programming approach where mathematical operations and variables are represented as draggable, interlocking blocks (Resnick et al. (2009)). This format reduces syntactic barriers, allowing learners to focus on the procedural logic of problem-solving. In mathematics education, block coding has been shown to enhance procedural fluency by enabling students to visualize and iteratively test computational steps without the cognitive load of conventional programming syntax. In addition, block coding has been shown to improve procedural thinking and facilitate the mapping between mathematical algorithms and executable code (Arslan Namli and Aybek (2022); Perin et al. (2023)). In this study, block coding serves three purposes: it facilitates procedural visualization by translating algebraic operations into step-by-step executable blocks, reinforces the connection between symbolic quadratic equation solving and algorithmic execution to bridge symbolic and computational thinking, and provides scaffolded automation by enabling a repeatable computational process to solve similar problems with different input values.

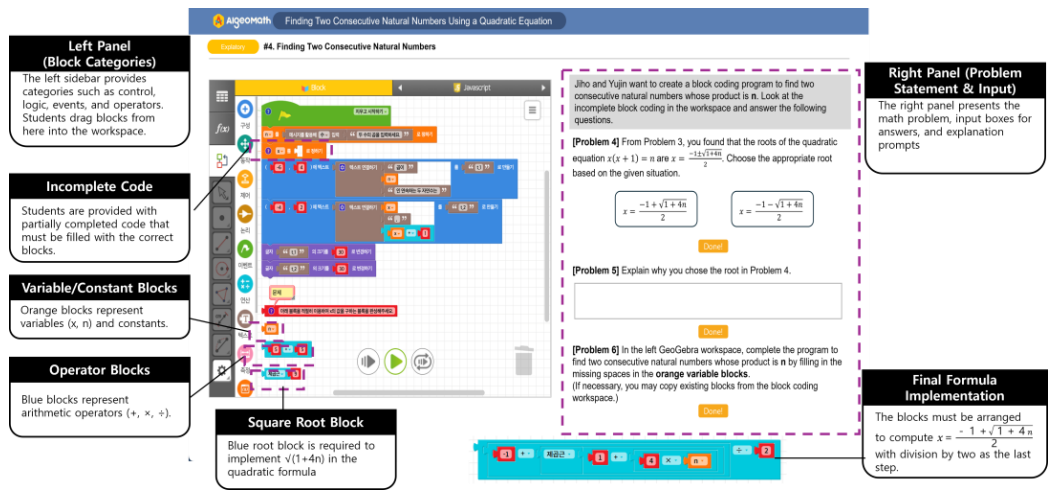

Figure 1: Example of the Algeomath block coding workspace used in Stages 4 and 5. Students complete the program to solve $x(x+1) = n$ using variable and operator blocks and implement $x = \frac{-1+\sqrt{1+4n}}{2}$ with division by two as the final operation.

In this stage, students: (i) explore how blocks execute step by step, (ii) reconstruct a given block sequence to compute simple expressions such as $a(b+c)$, and (iii) interpret the computational output as an algebraic expression.

**Stage #4** integrates symbolic reasoning with computational implementation. Students receive a partially completed block coding program designed to compute two consecutive natural numbers whose product is $n$ in an easy case. They must: (i) assign variables for the smaller and larger numbers, (ii) formulate the quadratic equation $x(x+1) = n$, (iii) solve for the roots and select the contextually appropriate solution, (iv) justify their choice, and (v) complete the missing code blocks. This stage explicitly links algebraic solution steps to their computational counterparts.

**Stage #5** repeats the Stage 4 procedure for a more difficult target value (e.g., 8742). The contrast between stages 4 and 5 highlights how computa-tional tools improve efficiency and accuracy when symbolic methods alone are time-consuming or error-prone. Figure 1 illustrates an example workspace where students complete the program implementing the quadratic formula $x = \frac{-1 + \sqrt{1 + 4n}}{2}$, with division by two as the final operation.

Table 3: Self-check survey items and reflection question

| # | Survey item (5-point Likert scale: Strongly agree – Strongly disagree) |
|---|---|
| 1 | I understand the meaning of a quadratic equation and its solutions. |
| 2 | I can solve a quadratic equation and explain the solution process. |
| 3 | I can represent the process of solving a quadratic equation using block coding. |
| 4 | I persist in solving real-world problems through mathematical modeling. |
| 5 | I recognize the usefulness of engineering tools such as block coding in mathematical problem-solving and can use them proactively. |
| **One-sentence reflection** | |
| Fill in the blanks: "Using ______, I learned ______, and in this process, I felt ______." | |

**Stage #6** shifts the focus from problem solving to meta-cognitive explanation. Students are asked to explain the relationship between values obtained through block coding and algebraic solutions. For example, they compare different implementations for consecutive even numbers, analyzing how variable definitions influence the computational outcome.

Finally, a **Self-check survey** was administered upon completion of all stages. As shown in Table 3, the survey consisted of five Likert-scale items aligned with the rubric (Table 2) and one open-ended reflection item. These responses were incorporated as supplementary evidence in the final academic achievement evaluation, consistent with prior findings that self-assessment enhances metacognitive awareness and performance appraisal (Köppe et al. (2024)).

*2.1.3. Web-Based Platform Implementation*

This study was implemented within a web-based environment built on AlgeomathKorea Foundation for the Advancement of Science and Creativity (KOFAC) (2023), a Korean online tool designed for interactive math education. The interface is divided into two panels: the left panel contains the block coding workspace, and the right panel presents the problem statement and input fields for student responses. Figure 1 illustrates an example of the block coding workspace. At each stage, students are allowed up to four submission attempts, receiving immediate grading and feedback after each attempt.

Grading in this system is tailored to the task type. For block coding tasks within the Algeomath workspace, the platform records a serialized representation of the program structure and parameters. This is matched against stage-specific reference solutions to verify both numerical correctness and adherence to required structural elements. For closed-ended answers, the server performs rule-based matching against a predefined set of correct responses, ensuring fast and deterministic grading. For open-ended explanations, the system uses few-shot prompting with an LLM, comparing the student's reasoning to expert-authored exemplars. The LLM outputs a structured correctness judgment, enabling flexible evaluation while maintaining alignment with instructional standards.

Feedback generation is based on a set of expert-authored error patterns and explanation templates. The system adapts the specificity of guidance according to the number of remaining attempts. In the first and second attempts, feedback emphasizes conceptual hints and expression-level refinements, while in the third and fourth attempts, it focuses on pinpointing key errors in equation formulation and block arrangement, along with corrective instructions. This progressive feedback structure is designed to maintain engagement and guide the learner toward successful problem completion, aligning with formative assessment principles.

All submission records and feedback messages are stored in a database, submitted answers, grading outcomes, and feedback text. These logs serve as input data for the final academic achievement evaluation, supporting rubric-based judgments across domains in Table 2.

The technical architecture, including communication protocols, prompt construction, and integration of the grading pipeline, is described in detail in (Lee et al. (2025)).

*2.2. Segment-Based Evidence Modeling for Rubric-Aligned Assessment*

*2.2.1. Segment Definition and Mapping to Rubric Subcategories*

In the proposed platform, each learning task is decomposed into *segments*, which correspond to the smallest independently evaluable units within a problem-solving sequence. A segment aligns with a single question in the learner's progression toward solving the overall task and is associated with a specific stage in the scenario map (Figure 2). The scenario map visualizes how segments are distributed across the three rubric domains — *Knowledge and Understanding*, *Procedural Skills*, and *Values and Attitudes* — and their respective subcategories. The *Procedural Skills* domain, which appears as a single category in Table 1, is subdivided in the scenario map into two distinct subcategories: solving quadratic equations (**Rublic 2**)and solving quadratic equations using block-based programming(**Rublic 3**). Similarly, the *Values and Attitudes* domain is split into two subcategories: a challenging attitude toward problem solving(**Rublic 4**), and proactive use of engineering tools(**Rublic 5**). In contrast, the *Knowledge and Understanding* domain remains as a single subcategory, that is Meaning of the Solutions of Quadratic Equations , focusing on understanding the solutions of quadratic equations (**Rublic 1**). This structure enables fine-grained evidence collection at the segment level, allowing multiple segments to contribute evidence toward the same rubric subcategory, and a single segment to support multiple subcategories if the task content overlaps.

By structuring evidence in this way, the system ensures that rubric-aligned evaluation is traceable to concrete learner actions. Figure 2 illustrates how the instructional flow is broken down into segments and mapped to rubric subcategories, establishing the many-to-many relationships that support holistic synthesis at the rubric level.

*2.2.2. Data Model and Evidence Aggregation for Rubric Aligned Evaluation*

The platform stores evidence at the segment level and then synthesizes it at the rubric level. A submission is saved in submission with the fields answers for the raw student response, answer_status for the correctness state, attempts for the number of tries, and a reference to the linked system feedback. When Algeomath is used, the computational work is captured as XML in algeo_answer. The XML encodes objects and parameters present in the workspace. During grading, the engine parses this XML against a stage specific template and extracts only the values introduced or edited by the learner, which are the parameters that operationalize the mathematical procedure. These extracted values and the numerical output are compared with the stage reference to decide answer_status. The result and a short rationale are stored in segment_evaluation and are tied to the submission identifier.

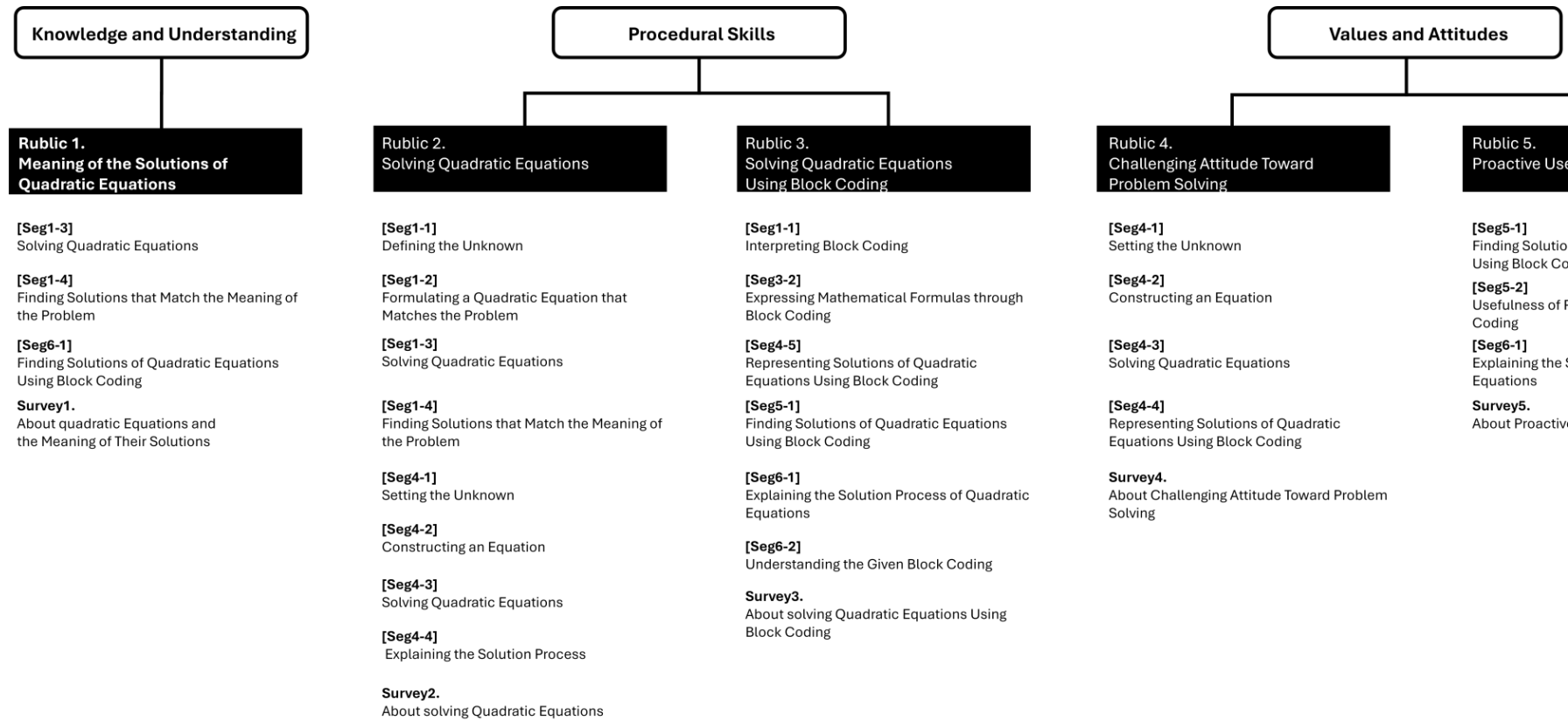

Figure 2: Scenario map showing the mapping of segments to rubric domains and subcategories. Segments, denoted as **[Seg X-Y]**, represent distinct evaluation units aligned to learning stages, and are connected to rubric subcategories within the domains of Knowledge and Understanding, Procedural Skills, and Values and Attitudes.

Rubric definitions reside in rubric and contain textual descriptors for the domain and the three performance levels. Each question carries a pointer to the relevant rubric entry so that multiple segments can supply evidence to the same sub category and a single segment can support several sub categories. This realises the many to many relationship visualised in Figure 2. Upon completion of all segments, the learner completes the self check recorded in lesson_evaluation, which contributes auxiliary evidence to the holistic judgment.

For holistic synthesis, the system gathers for each rubric sub category the ordered set of segments and their latest submissions together with any segment_evaluation and system feedback. Rubric text is included so that the model can anchor its judgment to explicit criteria. The self check score and the one sentence reflection are appended as auxiliary signals. These elements form the prompt to LLM under a constrained output schema. The model returns a categorical level among High, Medium, and Low and a concise rationale that cites specific segments. The outcome is written to rubric_evaluation with a session identifier so that every sentence can be traced back to concrete submissions and feedback messages.

These components are integrated into a complete pipeline, as shown in Figure 3. The resulting outputs are presented to learners and instructors through an interface (Figure 4) that summarizes both overall and rubric-based evaluations.

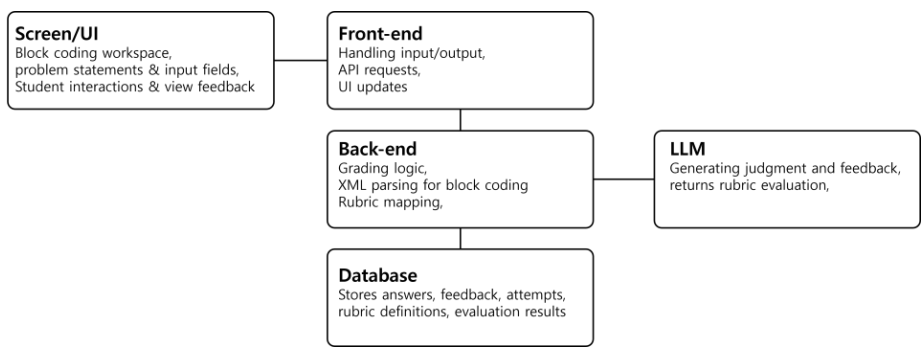

Figure 3: Overall system architecture: the client delivers tasks, the FastAPI backend grades and assembles evidence, the LLM provides rubric-aligned judgments, and PostgreSQL stores results.

## Overall Assessment

**1. Content of Evaluation**

The student has a basic understanding of solving quadratic equations but lacks detailed procedural and logical connections in the problem-solving process. There is a lack of clarity regarding the role of block coding and the function of each block.

**2. Evaluation Result**

*Medium*: The student can solve quadratic equations but lacks systematic approaches, preventing higher evaluation.

**3. Recommendations for Improvement**

The student needs to organize the solving procedure of quadratic equations step-by-step and clearly understand the role of each block. Studying various cases and actively adjusting personal approaches is encouraged.

**Overall Score: 59**

## Rubric Detailed Evaluations

| Rubric | Score | Self-Evaluation | Result | Recommendations Summary |
|---|---|---|---|---|
| Meaning of the Solutions of Quadratic Equations | 56 | STRONGLY_AGREE | Low | Clarify concepts and problem requirements; practice to reduce mistakes. |
| Solving Quadratic Equations | 81 | AGREE | Medium | Practice solution steps and conditions; repeat various problems. |
| Solving Quadratic Equations Using Block-Based Programming | 50 | DISAGREE | Low | Practice block connection and procedures; practice examples repeatedly. |
| A Challenging Attitude Toward Problem Solving | 75 | AGREE | Medium | Develop active use of tools and experience applying real cases. |
| Proactive Use of Engineering Tools | 33 | DISAGREE | Low | Study specific cases and concretize advantages and personal experiences. |

Figure 4: Summary of student assessment combining overall evaluation and detailed rubric results. The top section presents evaluation content, outcomes, and recommendations, while the table below reports rubric scores, self-evaluations, results, and improvement advice. For clearer presentation in the paper, the original multi-page web interface was condensed and edited into a single summary page.

## 3. Results

*3.1. Field Study Overview*

To examine the practical effectiveness of the rubric-based evaluation framework in classroom settings, we conducted a field study during November–December 2024 with 42 middle school students from two schools. All participants had previously learned quadratic equations and the quadratic formula, although prior exposure to block-based programming varied.

The study was implemented as a guided, technology-enhanced session. Teachers introduced the learning goals and platform usage, while a teaching assistant provided technical support. Students solved the multi-stage problems sequentially and were allowed up to four submission attempts per item. Submissions were automatically graded with real-time formative feedback, and all responses, attempts, and feedback logs were stored in the database for later analysis.

In addition to correctness checking, rubric-based evaluations were generated to capture multiple dimensions of performance, synthesized into an overall achievement score with descriptive commentary. Each student also compared the system's evaluation with their self-assessment, promoting reflective awareness of their problem-solving approach. The resulting dataset combines rubric-level evaluations, overall achievement measures, and detailed problem-solving histories, forming the basis for subsequent analyses of student achievement distributions, rubric-level performance, learner clusters, and the validity of LLM-generated feedback.

*3.2. Overall Performance Score Distribution Analysis*

For a comprehensive overview of students' overall academic achievement levels, the overall performance scores of 42 students were examined to describe the distribution characteristics, central tendency, and variability. The analysis was visualized using a histogram with a density curve to depict frequency patterns and a boxplot to highlight dispersion and potential outliers, as shown in Figure 5.

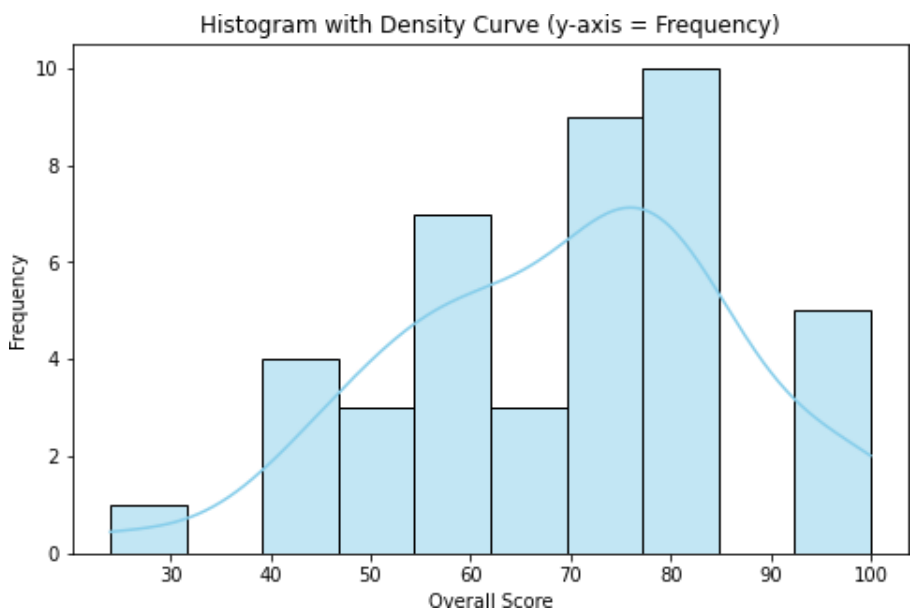

Figure 5: Overall performance score distribution. The histogram with density curve illustrates the frequency of scores across students.

The results indicate that the mean score was 69.81 with a standard deviation of 17.03, suggesting a relatively high degree of variation among students. Quartile analysis revealed that the lower 25% scored 59.0 or below, while the upper 25% scored 78.75 or above. The median score was 74.0. A few students scored as low as 24, representing very low achievement, whereas others reached the maximum of 100, reflecting outstanding performance. The histogram showed the highest concentration of students in the 70–80 range, while the boxplot identified both lower and upper extreme values.

Overall, these findings suggest that although the majority of students achieved scores near or above the intermediate level, the presence of both a lower-achieving group that may require targeted support and a higher-achieving group suitable for enrichment opportunities is evident.

*3.3. Rubric-wise Achievement Analysis*

To evaluate performance across different learning objectives, we extracted scores for Rubrics 1–5 from the detailed learning report. The average scores for each rubric revealed noticeable variation in achievement levels. Specifically, Rubrics 1 and 2 both recorded mean scores close to 80 points, indicating that students demonstrated strong understanding of the corresponding conceptual objectives. Rubric 4 also showed a similarly high level of performance, suggesting that students were relatively confident in applying their knowledge in structured contexts.

In contrast, Rubric 3 yielded a lower average score of approximately 65 points, reflecting moderate difficulty in problem-solving tasks that required deeper integration of learned concepts. The lowest performance was observed in Rubric 5, with an average score near 50 points and a comparatively wide standard deviation. This result highlights a particular weakness in more advanced or self-directed tasks, suggesting that additional scaffolding or targeted instructional support is needed in this area.

Figure 6 provides a visual summary of rubric-specific performance. The bar chart (left) displays mean scores with standard deviations for Rubrics 1–5, making clear the relative differences across objectives. The radar chart (right) further illustrates the performance profile, showing that while students' strengths lie in conceptual knowledge and structured applications (Rubrics 1, 2, and 4), there are consistent weaknesses in computational and higher-order applications (Rubrics 3 and 5). Together, these complementary visualizations highlight areas for both reinforcement and enrichment in instructional design.

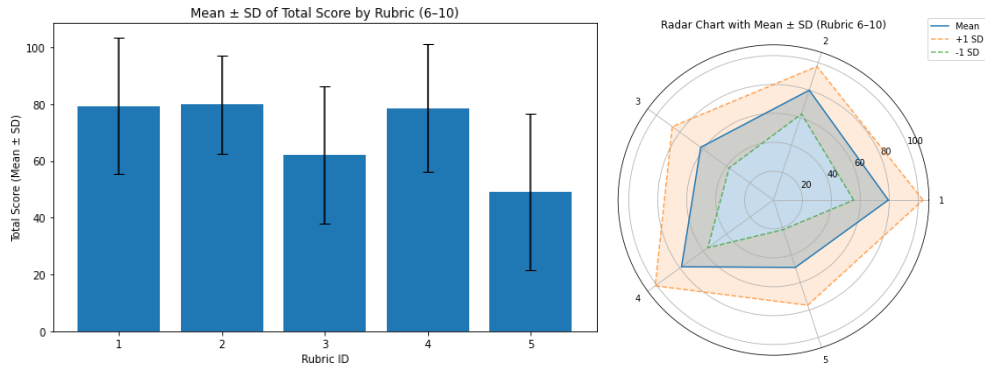

Figure 6: Rubric-wise achievement analysis. The bar chart (left) shows mean scores with standard deviations for Rubrics 1–5, while the radar chart (right) highlights relative strengths and weaknesses across rubrics.

*3.4. Learner Pattern and Cluster Analysis*

To explore heterogeneity in student performance profiles, a clustering analysis was conducted using rubric scores(Rubrics 1-5). Each rubric represents a distinct competency area, and the combined rubric score vector of each student was used as input for a K-means clustering algorithm. Based on silhouette analysis, the optimal number of clusters was determined to be $k = 6$. All rubric scores were standardized before clustering, and principal component analysis (PCA) was applied to reduce the five-dimensional rubric space into two dimensions for visualization.

The PCA scatter plot in Figure 7 shows the six clusters projected onto the first two principal components. While the first principal component (PC1) captures overall achievement variation, the second component (PC2) primarily reflects differences in rubric-specific performance patterns. The visualization indicates that students were separated into several small clusters, with some groups concentrated around high overall performance and others spread across lower or more unbalanced profiles.

To further interpret the clusters, Figure 7 presents rubric-wise achievement distributions for each cluster using violin plots. The comparison reveals distinct learner profiles. For example, some clusters showed consistently high performance in Rubric 1 (understanding the meaning of solutions) and Rubric 2 (solving quadratic equations), while exhibiting greater variability in Rubric 3 (block-based programming). Other clusters achieved relatively strong results in Rubric 4 (attitude toward problem solving) but displayed weaker outcomes in Rubric 5 (proactive use of engineering tools), suggesting that students were less confident in applying technological supports. A few clusters contained students with uniformly low performance across most rubrics, forming outlier-like groups that may require additional instructional support.

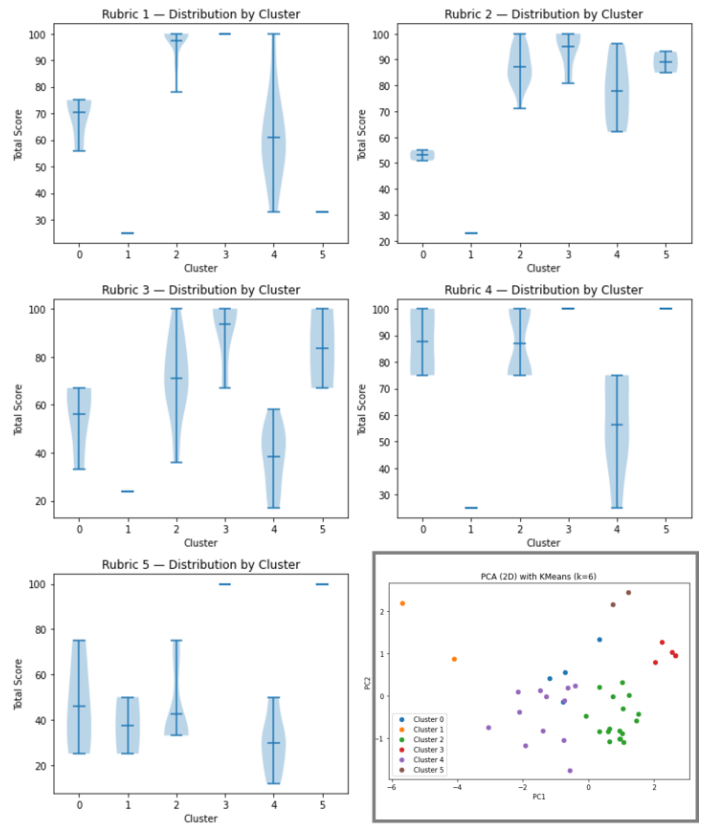

Figure 7: Clustering analysis of student performance. Violin plots (Rubrics 1–5) show rubric-wise score distributions by cluster, while the PCA scatter plot (bottom right) visualizes the six clusters ($k = 6$) projected onto the first two principal components.

These findings indicated that rubric-based evaluation not only differentiates students by overall score but also captures meaningful subgroups with unique competency profiles. In particular, weaknesses in Rubric 5 emerged as a common feature across multiple clusters, pointing to the need for targeted interventions in tasks involving higher-order application, self-directed work, or leadership in block coding activities.

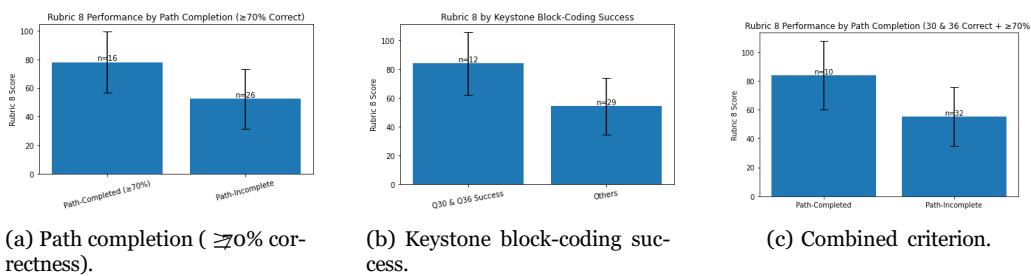

(a) Path completion (≥70% correctness).

(b) Keystone block-coding success.

(c) Combined criterion.

Figure 8: Rubric 3 score comparisons under different path-completion definitions: (a) overall scenario completion, (b) keystone block-coding success, and (c) combined criterion (path completion and keystone success).

*3.5. Scenario Path Completion and Competency in Block Coding*

To examine how different pathway criteria reflect learners' competency in block coding for quadratic equations, we compared Rubric 3 outcomes across three levels of scenario engagement.

First, using a simple threshold of 70% correctness across scenario tasks, students were classified into *path-completed* ($n = 16$) and *path-incomplete* ($n = 26$) groups. As shown in Figure 8a, the path-completed group achieved substantially higher Rubric 3 scores ($M = 78.8$, $SD = 18.7$) than the path-incomplete group ($M = 52.5$, $SD = 21.3$), with Welch's *t*-test confirming the significance of the difference ($t = 4.28$, $p < 0.001$).

Second, we focused on keystone block-coding items (Quesions realted [Seg 3-2] and [Seg 5-1]), which directly assess applied programming competency. Students who solved both items successfully ($n = 12$) attained significantly higher Rubric 3 scores ($M = 85.0$, $SD = 23.97$) compared to others ($n = 29$, $M = 55.3$, $SD = 20.32$), as shown in Figure 8b ($t = 3.42$, $p = 0.0045$). This result underscores the critical role of keystone tasks in distinguishing levels of block-coding proficiency.

Finally, we applied a stricter criterion that required both ≥70% overall scenario success and correct solutions to the keystone tasks (Quesions realted [Seg 3-2] and [Seg 5-1]). Under this combined definition, the path-completed group ($n = 10$) achieved the highest Rubric 3 mean score ($M = 84.0$, $SD = 23.97$), whereas the remainder ($n = 32$) averaged substantially lower ($M = 55.3$, $SD = 20.32$), as shown in Figure 8c. Welch's *t*-test again confirmed the robustness of the difference ($t = 3.42$, $p = 0.0045$).

These results mean that while broad scenario mastery (70% criterion) reflects general learning achievement, keystone items pinpoint essential com-petencies, and their combination provides the most stringent and pedagogically meaningful indicator of block-coding proficiency. This layered analysis underscores the importance of incorporating both general and keystone checkpoints into inquiry-based learning designs to promote comprehensive mastery.

*3.6. Alignment Between LLM-Derived Rubric Scores and Expert Judgments*

To assess the validity of the LLM-based rubric scoring, we compared its outcomes with expert-assigned ratings across five rubric dimensions. At the learner level, paired comparisons indicated that LLM-derived scores were closely aligned with expert evaluations: the average bias was modest (M=5.1 points), and error metrics remained within acceptable bounds (MAE = 12.4, RMSE = 15.8). Both Pearson and Spearman correlations confirmed strong positive associations between the two evaluation methods (r=0.79, $\rho$=0.76, both p<0.001), underscoring convergent validity. Complementary paired *t*-tests and Wilcoxon signed-rank tests showed no evidence of systematic divergence beyond this small upward bias in LLM scores.

Figure 9a illustrates the scatter alignment of expert versus LLM scores, with most points clustering along the 45-degree line. Bland–Altman analysis, shown in Figure 9b, further confirmed that differences were symmetrically distributed, with limits of agreement (± 1.96 standard deviation) capturing the majority of cases without extreme outliers. The histogram of differences (Figure 9c) highlights that the modal discrepancy was close to zero, while rubric-wise violin plots, shown in Figure 9d reveal comparable alignment across all rubric categories, albeit with slightly wider dispersion in Rubric 4.

These results indicate that the LLM-driven rubric assessment provides a reliable approximation of expert judgments, not only reproducing overall score distributions but also preserving rubric-level consistency. This alignment suggests that LLM scoring can serve as a practical proxy for expert evaluation in formative assessment contexts, reducing the resource burden of manual scoring while maintaining interpretive validity.

## 4. Conclusion

This study proposed a rubric-aligned assessment framework grounded in segment-based evidence modeling. By decomposing multi-step mathematics problem-solving tasks into segments aligned with rubric subcategories across Knowledge and Understanding, Procedural Skills, and Values and Attitudes,

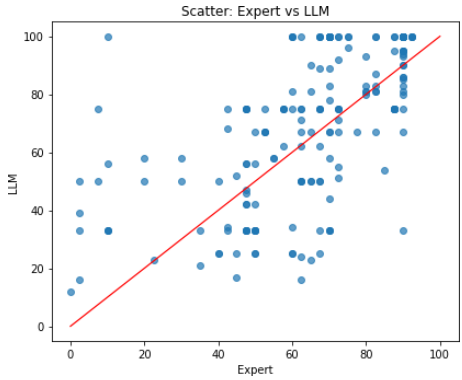

(a) Scatter plot comparing expert and LLM rubric scores. The red line indicates perfect agreement.

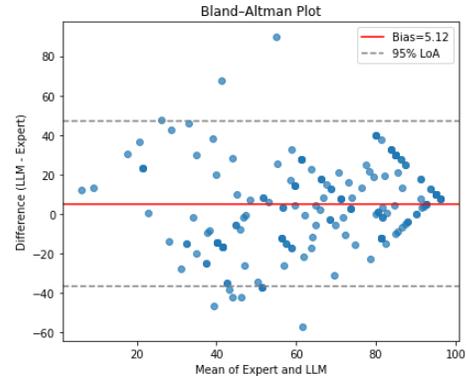

(b) Bland–Altman plot of LLM versus expert scores. The red line represents the mean bias and dashed lines indicate 95% limits of agreement.

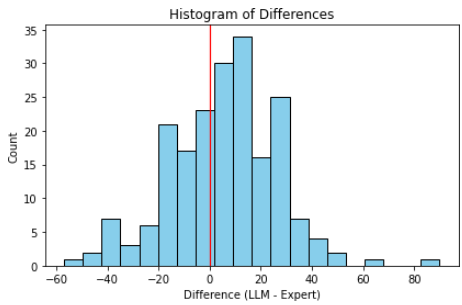

(c) Histogram of score differences (LLM− Expert). The distribution centers near zero, indicating minimal systematic bias.

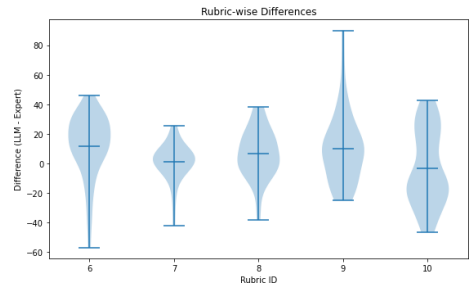

(d) Rubric-wise violin plots of LLM–Expert score differences (Rubrics 1–5). Rubric 4 shows slightly larger variability.

Figure 9: Comparison of LLM-derived rubric scores with expert evaluations: (a) scatter plot, (b) Bland–Altman plot, (c) histogram of differences, and (d) rubric-wise violin plots.

the system ensured that every rubric judgment could be transparently traced back to concrete learner actions. The many-to-many mapping between segments and rubric subcategories enabled evidence to be aggregated holistically while maintaining interpretability, allowing the LLM to generate judgments that were both criterion-referenced and process-aware.

We implemented and field-tested a web-based mathematics learning platform that integrates automated grading and feedback with rubric-aligned achievement evaluation. By combining rule-based grading for objective responses and LLM-assisted synthesis for open-ended and process-oriented tasks, the system provided both fine-grained segment-level assessment and holistic rubric-based judgments.

Field deployment with 42 middle-school students demonstrated that the platform successfully captured the full trajectory of problem solving, including intermediate attempts and revisions, thereby enabling a process-centered evaluation that conventional assessment methods often fail to achieve. The LLM's capacity to aggregate and interpret diverse forms of unstructured data (answers, iterative feedback, self-assessments) yielded coherent evaluations that were pedagogically meaningful. Importantly, mathematics education experts noted that tracing learning trajectories across multiple attempts provided valuable diagnostic insight, while the LLM-generated synthesis substantially reduced the cognitive and temporal burden of manual grading.

Although this study was conducted with a limited dataset and restricted to quadratic equation problems, the results indicate that LLM-based rubric scoring can be a viable solution for automated, process-sensitive assessment. Future work will aim to validate these findings with larger datasets, apply the approach to a broader range of mathematical and STEM problems, and systematically investigate how prompt design and rubric calibration impact evaluation accuracy. Ultimately, this work lays the groundwork for building more scalable and robust assessment systems for educational applications.

## Acknowledgment

We thank Jongsuk Yoon and Joohyun Lee for their valuable assistance with the implementation of the system. Their support and technical contributions were instrumental to this work. This research was supported in 2024 by the Korea Foundation for the Advancement of Science and Creativity (KOFAC) under the project titled "Development of an Integrated Engineering Tool Model and Prototype for Inquiry-Activity-Based Assessment." The study was reviewed and approved by the Institutional Review Board of Hongik University (IRB No. 7002340-202409-HR021, approval date: September 3, 2024) and conducted in accordance with the principles outlined in the Declaration of Helsinki.